\documentclass[twocolumn,preprintnumbers,a4paper,
superscriptaddress,floatfix,twoside,prd]{revtex4}

\usepackage{bm}
\usepackage{epsfig}
\usepackage{graphics}
\usepackage{natbib}

\usepackage{fancyh}
\usepackage[l]{floatflt}
\usepackage{epsfig}
\usepackage{amssymb}
\usepackage{latexsym}
\usepackage{times}
\usepackage{slashed}
\usepackage{upgreek}
\usepackage{amsmath}
\usepackage{amsfonts}
\usepackage{amsbsy}
\usepackage{amscd}
\usepackage{bbm}

\newcommand{\Eref}[1]{Eq.~(\ref{#1})}

\newcommand{\Sref}[1]{Sec.~\ref{#1}}
\newcommand{\Fref}[1]{Fig.~\ref{#1}}
\newcommand{\Tref}[1]{Table~\ref{#1}}
\newcommand{\cref}[1]{Ref.~\cite{#1}}
\newcommand{\crefs}[1]{Refs.~\cite{#1}}

\newcommand{\hepph}[1]{{\ftn\tt hep-ph/#1}}

\newcommand{\astroph}[1]{{\ftn\tt astro-ph/#1}}
\newcommand{\arxiv}[1]{{\ftn\tt  arXiv:#1}}

\newcommand{\bal}{\begin{align}}
\newcommand{\eall}{\end{align}}
\newcommand{\beqs}{\begin{subequations}}
\newcommand{\eeqs}{\end{subequations}}
\newcommand{\eec}{\end{center}}
\newcommand{\bec}{\begin{center}}

\newcommand{\eem}{\end{matrix}}
\newcommand{\bem}{\begin{matrix}}
\newcommand{\eeq}{\end{equation}}
\newcommand{\beq}{\begin{equation}}
\newcommand{\ba}{\begin{array}}
\newcommand{\ea}{\end{array}}
\newcommand{\bea}{\begin{eqnarray}}
\newcommand{\eea}{\end{eqnarray}}
\newcommand{\baq}{\begin{eqnarray}}
\newcommand{\eaq}{\end{eqnarray}}
\newcommand{\bel}{\begin{align}}
\newcommand{\eeel}{\begin{align}}

\newcommand{\beld}{\begin{aligned}}
\newcommand{\eeld}{\begin{aligned}}

\newcommand\eqs[2]{Eqs.~(\ref{#1}) and (\ref{#2})}

\newcommand{\ftn}{\footnotesize}

\newcommand{\sEref}[2]{Eq.~(\ref{#1}{\small\sf {#2}})}

\newcommand{\etal}{{\it et al.\/}}

\def\to{\rightarrow}

\def\llgm{\left\lgroup}
\def\rrgm{\right\rgroup}
\def\lf{\left(}
\def\rg{\right)}
\newcommand\vev[1]{\left\langle {#1} \right\rangle}

\newcommand{\Vhi}{\ensuremath{\widehat V_{\rm CI}}}
\newcommand{\Vci}{\ensuremath{\widehat V_{\rm CI}}}
\newcommand{\Vjhi}{\ensuremath{V_{\rm CI}}}
\newcommand{\Hhi}{\ensuremath{\widehat H_{\rm CI}}}

\newcommand{\Khi}{\ensuremath{K}}
\newcommand{\Whi}{\ensuremath{W}}

\newcommand{\Ns}{\ensuremath{{\what N_\star}}}

\newcommand{\mP}{\ensuremath{m_{\rm P}}}

\newcommand{\Qef}{\ensuremath{\Lambda_{\rm UV}}}

\def\openone{\leavevmode\hbox{\small1\kern-3.8pt\normalsize1}}
\newcommand{\dV}{\ensuremath{\Delta\widehat V_{\rm CI}}}

\newcommand{\fr}{\ensuremath{f_{\cal R}}}

\newcommand{\cat}{\ensuremath{c_{\cal R}}}
\newcommand{\ca}{\ensuremath{c_{1\cal R}}}
\newcommand{\cb}{\ensuremath{c_{2\cal R}}}

\newcommand{\ks}{\ensuremath{k_\star}}
\newcommand{\ns}{\ensuremath{n_{\rm s}}}

\newcommand{\as}{\ensuremath{a_{\rm s}}}
\newcommand{\As}{\ensuremath{A_{\rm s}}}
\newcommand{\rw}{\ensuremath{r_{0.002}}}
\newcommand{\rs}{\ensuremath{r_{21}}}
\newcommand{\rcc}{\ensuremath{\mathcal{R}}}
\newcommand{\rce}{\ensuremath{\widehat{\mathcal{R}}}}
\newcommand{\Ve}{\ensuremath{\widehat{V}}}

\newcommand{\ern}{\ensuremath{e_n}}

\newcommand{\dphi}{\ensuremath{\what{\delta\phi}}}
\newcommand{\dph}{\ensuremath{\delta\phi}}

\newcommand{\fm}{\ensuremath{F_{-}}}
\newcommand{\fp}{\ensuremath{F_{\cal R}}}

\newcommand\mtta[4]{\mbox{
$\llgm\bem #1 &#2 \cr #3& #4\eem\rrgm$}}

\newcommand{\what}{\ensuremath{\widehat}}

\def\aal{{\bar\alpha}}
\def\bbet{{\bar\beta}}
\def\al{{\alpha}}

\def\K{{\widehat{K}}}

\def\n{\bar{n}}

\def\th{{\theta}}

\newcommand{\sg}{\ensuremath{\phi}}

\newcommand{\sgx}{\ensuremath{\phi_\star}}
\newcommand{\sgf}{\ensuremath{\phi_{\rm f}}}

\newcommand{\ld}{\ensuremath{\lambda}}

\newcommand{\Ld}{\ensuremath{\Lambda}}

\newcommand{\se}{\ensuremath{\widehat \phi}}
\newcommand{\sex}{\ensuremath{\widehat{\phi}_\star}}
\newcommand{\sef}{\ensuremath{\widehat{\phi}_{\rm f}}}
\newcommand{\geu}{\ensuremath{\widehat g}}
\newcommand{\eph}{\ensuremath{\widehat \epsilon}}
\newcommand{\ith}{\ensuremath{\widehat \eta}}

\newcommand{\phc}{\ensuremath{\Phi}}
\newcommand{\phcb}{\ensuremath{\bar\Phi}}

\def\Ka{K\"{a}hler potential}

\def\sub{subplanckian}

\def\bcp{{\sc\small Bicep2}/{\it Keck Array}}
\newcommand{\plk}{{\it Planck}}

\newcommand{\diag}{\ensuremath{{\sf diag}}}
\newcommand{\im}{\ensuremath{{\sf Im}}}

\renewcommand{\refname}{{\bf\scshape References}}

\renewcommand{\thesubsection}{{\small\sf\Alph{subsection}}}

\renewenvironment{subequations}{%
\refstepcounter{equation}%
\setcounter{parentequation}{\value{equation}}%
  \setcounter{equation}{0}
  \ignorespaces
}{%
  \setcounter{equation}{\value{parentequation}}%
  \ignorespacesafterend
}

\begin{document}


\title{\bf\scshape Unitarizing non-Minimal Inflation via \\ a Linear Contribution to the Frame Function}

\author{\scshape Constantinos Pallis\\ {\it School of Technology,
Aristotle University of Thessaloniki, GR-541 12 Thessaloniki,
GREECE}\\  {\sl e-mail address: }{\ftn\tt kpallis@gen.auth.gr}}



\begin{abstract}

\noindent {\ftn \bf\scshape Abstract:} We show that non-minimal
inflation, based on the $\sg^4$ potential, may be rendered
unitarity conserving and compatible with the \plk\ results for
$4.6\cdot10^{-3}\leq\rs=\cb/\ca^{2}\leq1$, if we introduce a
linear contribution ($\ca\phi$) to the frame function which takes
the form $\fr=1+\ca\phi+\cb\phi^2$. Supersymmetrization of this
model can be achieved by considering two gauge singlet superfields
and combining a linear-quadratic superpotential term with a class
of logarithmic or semi-logarithmic \Ka s with prefactor for the
logarithms including the inflaton field $-(2n+3)$ or $-2(n+1)$
where $-0.01\lesssim n\lesssim0.013$.
\\ \\ {\scriptsize {\sf PACs numbers: 98.80.Cq, 04.50.Kd, 04.65.+e}
\hfill {\sl\bfseries Published in} {\sl Phys. Lett B} {\bf 789},
243 (2019)}

\end{abstract}\pagestyle{fancyplain}

\maketitle

\rhead[\fancyplain{}{ \bf \thepage}]{\fancyplain{}{\sl Unitarizing
nMI via a Linear Contribution to the Frame Function}}
\lhead[\fancyplain{}{\sl C. Pallis}]{\fancyplain{}{\bf \thepage}}
\cfoot{}

\section{Introduction}\label{intro}

Although excellently compatible with data \cite{plcp, plin,
gwsnew}, (quartic) \emph{non-minimal inflation} ({\sf\ftn nMI})
based on the potential
\beq \label{Vn}\Vjhi(\sg)=\ld^2\sg^4/4\eeq
and realized \cite{old,sm1} thanks to the presence of a
non-minimal coupling function
\beq \label{frattr} \fr(\sg)=1+\cat\sg^{2}  \eeq
between the inflaton $\sg$ and the Ricci scalar $\rcc$, suffers
from an inconsistency \cite{cutoff, riotto} with the validity of
the effective theory. Indeed, the establishment of the
inflationary stage with $\sg\leq1$ -- in the reduced Planck units
with $\mP=1$ -- requires large $\cat$ values which drive the
\emph{Einstein frame} ({\sf\ftn EF}) inflationary scale \beq\what
V_{\rm CI}^{1/4}=\Vjhi^{1/4}/\fr^{1/2} \label{Vhiattr}\eeq to
values well above the \emph{Ultraviolet} ({\sf\ftn UV}) cut-off
scale
\beq\Qef=\mP/\cat\label{Qattr}\eeq
of the effective theory which, thereby, breaks down above it  -- a
criticism of these results may be found in \cref{cutof}. Several
ways have been proposed to surpass this inconsistency. E.g.,
incorporating new degrees of freedom at $\Qef$ \cite{gianlee}, or
assuming additional interactions \cite{john}, or introducing a
sizable kinetic mixing in the inflaton sector which dominates over
$\fr$ \cite{lee, nMkin, jhep, var} or even adding an $\rcc^2$ term
\cite{tokareva}. In the case of high-scale nMI this problem can be
also eluded invoking a large inflaton \emph{vacuum expectation
value} ({\sf\ftn v.e.v}) $\vev{\sg}$ as in \crefs{R2r, nIG, igHI,
gian}. Unitarity-conserving nMI can be also achieved by
considering a scalar field, which exhibits a linear contribution
to its non-minimal coupling to gravity, in addition to the
\emph{Standard Model} ({\sf\ftn SM}) Higgs field which is coupled
quadratically to gravity \cite{jose, leel}.

In a recent paper \cite{flinear} we propose a novel solution to
the aforementioned problem which works only in the context of
\emph{Supergravity} ({\sf\ftn SUGRA}) and is relied on the
adoption of an exclusively linear non-minimal coupling to gravity
in conjunction with an appropriate selection of the prefactor of
the logarithm of the \Ka. We here suggest an alternative solution,
operative also in non-SUSY settings. It is tied to the presence of
a subdominant linear term into $\fr$, which thereby takes the form
\beq \fr(\sg)=1+\ca\sg+\cb\sg^2\,.\label{fr}\eeq
In a such case, the canonically normalized inflaton $\se$ is
related to the initial field $\sg$ as $\se\sim\ca\sg$ at the
vacuum of the theory, in sharp contrast to what we obtain for
$\fr$ in \Eref{frattr} where $\se\simeq\sg$. Indeed, $\se$ is
given in terms of $\sg$ using the formula \cite{sm1}
\beq \label{Je}
\frac{d\se}{d\sg}=\sqrt{\frac{1}{\fr}+{3\over2}\left({f_{\cal
R,\sg}\over \fr}\right)^2}\eeq
where the symbol $,\phi$ as subscript denotes derivation
\emph{with respect to} ({\ftn\sf w.r.t}) the field $\phi$. From
\Eref{Je} we can easily infer that if $\fr$ is linear
\cite{riotto,quad} or if it includes a linear contribution
$f_{{\cal R},\sg}\neq0$ and so $\se\neq\sg$ at the vacuum of the
theory which typically is given by the condition $\vev{\sg}=0$. As
a consequence, the small-field series of the various terms of the
action, expressed in terms of $\se$, contain powers of the ratio
$\rs=\cb/\ca^2$ -- and not of the parameters $\ca$ or $\cb$
appearing in the \emph{right-hand side} ({\sf\ftn r.h.s}) of
\Eref{fr} -- preventing, thereby, the reduction of $\Qef$ below
$\mP$ for $\rs\leq1$, despite the fact that $\ca$ and $\cb$ may be
large.

Although the present proposal is ``tailor-made'' for the non-SUSY
regime of the quartic nMI, we prefer to investigate the relevant
setting in the context of SUGRA in order to enrich the parameter
space of the model and highlight its differences with the proposal
of \cref{flinear}. Indeed, the emergent picture here is radically
different from that found in \cref{flinear}. Namely, the
inflationary potential is of Starobinsky type and the observables
crucially depend on the ratio $\rs$ which is an extra parameter
w.r.t those employed in \cref{flinear}. On the other hand, we do
not consider any mixing of the inflaton with other fields as in
\cref{leel} and so our setting is considerably simplified.

Below, in \Sref{setup}, we describe how we can formulate this kind
of unitarity-safe nMI both within a SUSY and non-SUSY framework.
The dynamics of the resulting inflationary models is studied in
\Sref{inf} and these are tested against observations in
\Sref{res}. Finally, we analyze the UV behavior of the models in
Secs.~\ref{uv} and summarize our conclusions  in \Sref{con}.

\section{SUSY Versus non-SUSY Framework}\label{setup}


Here we shortly remind the establishment of nMI within a non-SUSY
framework -- in \Sref{nonsusy} -- and then in the context of SUGRA
-- see \Sref{hi}.

\subsection{\small\sf  Non-SUSY Setting} \label{nonsusy}

Non-Minimal Inflation (i.e., nMI) is formulated in the
\emph{Jordan frame} ({\sf\ftn JF}) where the action of the
inflaton $\phi$ is given by
\beqs\beq \label{actionJ} {\sf  S} = \int d^4 x
\sqrt{-\mathfrak{g}} \left(-\frac{\fr}{2}\rcc +\frac12g^{\mu\nu}
\partial_\mu \sg\partial_\nu \sg-
\Vjhi(\sg)\right). \eeq
Here $\mathfrak{g}$ is the determinant of the background
Friedmann-Robertson-Walker metric, $g^{\mu\nu}$ with signature
$(+,-,-,-)$ whereas $V$ is given by \Eref{Vn}. By performing a
conformal transformation \cite{sm1} according to which we define
the EF metric $\geu_{\mu\nu}=\fr\,g_{\mu\nu}$ with determinant
$\widehat{\mathfrak{g}}$, we can write ${\sf S}$ as follows
\beq {\sf  S}= \int d^4 x
\sqrt{-\what{\mathfrak{g}}}\left(-\frac12
\rce+\frac12\geu^{\mu\nu} \partial_\mu \se\partial_\nu \se
-\Vhi(\se)\right), \label{action} \eeq\eeqs
where $\rce$ is the EF Ricci scalar curvature, $\Vhi$ is given as
a function of $\sg$ by the virtue of \Eref{Vhiattr} with $\fr$
defined by \Eref{fr}. If we wish to couple this model to the SM,
we have to assume that the SM fields are minimally coupled to
gravity and the potential mixing of $\sg$ to the SM Higgs field is
very weak -- cf.~\cref{leel}.


\subsection{\small\sf  Supergravity Embeddings} \label{hi}

A convenient implementation of nMI in SUGRA is achieved by
employing two singlet superfields $z^\al=\Phi, S$, with $\Phi$
($\al=1$) and $S$ ($\al=2)$ being the inflaton and a
``stabilizer'' field respectively. We below describe the salient
feature of our SUGRA setting in \Sref{hi1} and outline the
derivation of the inflationary potential in \Sref{hi2}.

\subsubsection{\small\sf Set-up}\label{hi1}

The EF action for $z^\al$'s  can be written as \cite{linde1}
\beq\label{action1} {\sf S}=\int d^4x \sqrt{-\what{
\mathfrak{g}}}\lf-\frac{1}{2} \rce +K_{\al\bbet}\geu^{\mu\nu}
\partial_\mu z^\al \partial_\nu z^{*\bbet}-\Ve\rg, \eeq
where the summation is taken over the scalar fields $z^\al$,
$K_{\al\bbet}={\K_{,z^\al z^{*\bbet}}}$ with
$K^{\bbet\al}K_{\al\bar \gamma}=\delta^\bbet_{\bar \gamma}$ and
$\Ve$ is the EF F--term SUGRA scalar potential which can be
extracted once the superpotential $\Whi$ and the \Ka\ $\Khi$ have
been selected, via the formula
\beq \Ve=e^{\Khi}\lf K^{\al\bbet}D_\al \Whi D^*_\bbet
\Whi^*-3{\vert \Whi\vert^2}\rg\label{Vsugra} \eeq
where $D_\al W=W_{,z^\al} +K_{,z^\al}W$ is the K\"ahler covariant
derivative.

The presence of the stabilizer field $S$ facilitates the
reproduction of \Eref{Vhiattr} from \Eref{Vsugra} by placing $S$
at the origin. Then, the only surviving term in \Eref{Vsugra} is
\beq \label{Vhi1}\Vhi=e^{K}K^{SS^*}\, |W_{,S}|^2\,,\eeq
and the numerator in \Eref{Vhiattr}, originating from \Eref{Vn},
can be derived if we adopt the following superpotential
\beq \Whi=\ld S\Phi^2\,.\label{Whi} \eeq
$\Whi$ can be uniquely determined if we impose two symmetries:
{\sf\ftn (i)} an $R$ symmetry under which $S$ and $\Phi$ have
charges $1$ and $0$; {\sf\ftn (ii)} a global $U(1)$ symmetry with
assigned charges $-1$ and $1$ for $S$ and $\Phi$. The derivation
of the denominator in \Eref{Vhiattr}, with $\fr$ defined in
\Eref{fr}, can be obtained, though, by violating the latter
symmetry as regards $\phc$. Indeed, we employ one of the \Ka s
below
\beqs\bea
K_1&=&-N\ln\left(1+\fp+\fp^*-\fm/N+F_{1S}\right),~~~\label{K1}\\
K_2&=&-N\ln\left(1+\fp+\fp^*-\fm/N\right)+F_{2S},\label{K2}
\eea\eeqs
where $N>0$ and the functions $\fp$ and $\fm$ are defined as
\beq \label{fpm}
\fp=\ca\phc/\sqrt{2}+\cb\phc^2~~~\mbox{and}~~~\fm=-\frac12\lf\phc-\phc^*\rg^2\,.\eeq
From these, the first one allows for the introduction of the
polyonimic non-minimal coupling $\fr$ in \Eref{fr}, whereas the
second one assures canonical normalization of $\Phi$ without any
contribution to the non-minimal coupling along the inflationary
path -- cf.~\cref{nMkin}. On the other hand, the functions
$F_{lS}$ with $l=1,2$ are defined as
\beqs\bea
F_{1S}&=&-\ln\left(1+|S|^2/N\right),\label{f1s}\\
F_{2S}&=&N_S\ln\left(1+|S|^2/N_S\right),\label{f2s} \eea\eeqs
and offer canonical normalization and safe stabilization of $S$
during and after nMI \cite{su11}. We avoid here to study $K$'s
obtained by placing $\fm$ outside the argument of $\ln$ since the
resulting models exhibit a more complicate inflationary dynamics
which leads to observables drastically deviating from those in
non-SUSY case, as shown in \cref{flinear}.

The construction of \Eref{actionJ} can be obtained within SUGRA if
we perform the inverse of the conformal transformation described
above \Eref{action} with
\beq \fr=-\Omega/N,\label{Omgfr}\eeq
and specify the following relation between $K$ and $\Omega$,
\beq-\Omega/N
=e^{-K/N}\>\Rightarrow\>K=-N\ln\lf-\Omega/N\rg\,.\label{Omg1}\eeq
Working along the lines of \crefs{linde1,quad} we arrive at the JF
action
\begin{equation}\begin{aligned}\label{Sfinal} {\sf S}&=\int d^4x
\sqrt{-\mathfrak{g}}\lf\frac{\Omega}{2N}\rcc+{\cal
\omega}_{\al\bbet}\partial_\mu z^\al \partial^\mu z^{*\bbet}-V
\right.\\ &- \left.\frac{27}{N^3}\Omega{\cal A}_\mu{\cal
A}^\mu\rg~~\mbox{with}~~{\cal
\omega}_{\al\bbet}=\Omega_{\al{\bbet}}+\frac{3-N}{N}\frac{\Omega_{\al}\Omega_{\bbet}}{\Omega}\,\cdot
\end{aligned}\end{equation}
Here $V =\Omega^2\Ve/N^2$ is the JF potential and ${\cal A}_\mu$
is \cite{linde1} the purely bosonic part of the on-shell value of
the auxiliary field found to be \beq {\cal A}_\mu =-iN\lf
\Omega_\al\partial_\mu z^\al-\Omega_\aal\partial_\mu
z^{*\aal}\rg/{6\Omega}.\eeq We note that, contrary to the non-SUSY
case -- see \Eref{actionJ} --, we obtain a kinetic mixing in
\Eref{Sfinal} which, along the inflationary trough $\im\phc=S=0$,
can be cast in the form
\beqs\beq \label{omab} {\cal \omega}_{\al\bbet}=
\mtta{1+(N-3)f_{{\cal R},\sg}^2/4\fr}{0}{0}{\fr K_{SS^*}},\eeq
where
\beq \label{Kss} K_{SS^*}=\begin{cases}
1/\fr&\mbox{for}~~K=K_1,\\1&\mbox{for}~~K=K_2\,.\end{cases}
\eeq\eeqs
From \Eref{omab} we see that canonical kinetic terms arise for
$N=3$ and $K=K_1$. On the contrary, for $N\neq3$ there is some
kinetic mixing which, however, does not disturb essentially the
realization of nMI and allows us to obtain adjustable inflationary
observables -- see \Sref{inf}.

\subsubsection{\sf\small Inflationary Potential}\label{hi2}

We here verify that the proposed $W$ and $K$'s in \eqs{Whi}{K1} or
(\ref{K2}) result to an inflationary model approaching $\Vhi$ in
\Eref{Vhiattr} with $\fr$ defined in \Eref{fr} and $\se$ found by
\Eref{Je}. Indeed, computing $\Vhi$ in \Eref{Vhi1}, along the
direction
\beq \label{inftr} \phc=\phc^*~~\mbox{and}~~S=0,~~\mbox{or}~~\bar
s=s=\th=0\,,\eeq if we express $\Phi$ and $S$ according to the
parametrization
\beq\label{hpar} \Phi={\sg e^{i\th}}/{\sqrt{2}}~~\mbox{and}~~S=(s
+i\bar s)/{\sqrt{2}}\,,\eeq
we can extract the final form of $\Vhi$ in \Eref{Vhi1}
\beq \label{Vhi}\Vhi
=\frac{\ld^2\sg^4}{4\fr^{2(n+1)}}=\frac{\ld^2\sg^4}{4\fr^{N}}\cdot\begin{cases}
\fr&\mbox{for}\>\>K=K_1\\
1&\mbox{for}\>\>K= K_2.\end{cases}\eeq
Here we take into account \Eref{Kss} and introduce $n$ through the
relation
\beq \label{ndef} N= \begin{cases} 2n+3 &\mbox{for}\>\> K=K_1, \\
2(n+1) &\mbox{for}\>\> K=K_2.
\end{cases}\eeq
For $n=0$, $\Vci$ reduces to the one obtained in \Eref{Vhiattr}
with $\fr$ shown in \Eref{fr}. This choice is special since it
yields integer $N$'s which are more friendly to string theory.
However, non-integer $N$'s are also acceptable \cite{roest, jhep,
var, nIG} and assist us to cover the whole allowed domain of the
observables. More specifically, for $n<0$, $\Vci$ remains an
increasing function of $\sg$, whereas for $n>0$, it develops a
local maximum $\Vci(\sg_{\rm max})$ where
\beq \sg_{\rm max}=\frac4{\ca}\lf n-1+\sqrt{(n-1)^2 + 16 n
\rs}\rg^{-1}\,.\label{Vmax}\eeq
In a such case we are forced to assume that hilltop \cite{lofti}
nMI occurs with $\sg$ rolling from the region of the maximum down
to smaller values.

To specify the EF canonically normalized inflaton, we note that,
for both $K$'s in \eqs{K1}{K2}, $K_{\al\bbet}$ along the
configuration in \Eref{inftr} takes the form
\beq \lf K_{\al\bbet}\rg=\diag\lf
K_{\phc\phc^*},K_{SS^*}\rg,\label{Kab} \eeq
where $K_{SS^*}$ is given by \Eref{Kss} and
\beq \label{kp}
K_{\phc\phcb}=(2\fr+N\ca^2+4N\cb^2\sg^2+4N\ca\cb\sg)/2\fr^2\,.\eeq
Therefore, the EF canonically normalized fields, denoted by hat,
are defined via the relations
\beq  \label{Jg} \frac{d\se}{d\sg}=\sqrt{K_{\Phi\Phi^*}}=J,\>\>
\what{\th}= J\th\sg\>\>\mbox{and}\>\>(\what s,\what{\bar
s})=\sqrt{K_{SS^*}} {(s,\bar s)}\eeq
and the spinors $\psi_S$ and $\psi_{\Phi}$ associated with the
superfields $S$ and $\Phi$ are normalized similarly, i.e.,
$\what\psi_{S}=\sqrt{K_{SS^*}}\psi_{S}$ and
$\what\psi_{\Phi}=J\psi_{\Phi}$. For $N=3$ the leftmost equality
in \Eref{Jg} reduces to \Eref{Je} which is valid in the non-SUSY
regime.

\begin{table}[t!]
\caption{\normalfont Mass squared spectrum along the path in
\Eref{inftr}.}
\begin{ruledtabular}
\begin{tabular}{c|c|c|c}
%
{\sc Fields}&{\sc Einge-} & \multicolumn{2}{c}{\sc Mass
Squared}\\\cline{3-4} &{\sc states}&$K=K_1$&$K=K_2$\\\hline
%
%
$1$ real scalar &$\what \th$ & $6(1-1/N)\Hhi^2$&$6\Hhi^2$\\
$2$ real scalars &$\what{s},\what{\bar s}$ & $6\cb\Hhi^2\sg^2/N$&$6\Hhi^2/N_S$\\
\hline\\[-0.4cm]
%
%
$2$ Weyl &$\frac{\what{\psi}_{S}\pm
\what{\psi}_{\Phi}}{\sqrt{2}}~~$&
\multicolumn{2}{c}{$(4-\ca(N-4)\sg-2\bar
N\cb\sg^2)^2$}\\spinors&&\multicolumn{2}{c}{$3\Hhi^2/4N\cb^2\sg^4$}\\\cline{3-4}
&&$\bar N=N-3$&$\bar N=N-2$
\end{tabular}
\end{ruledtabular}\label{tab1}
\end{table}

Taking the limit $\ca\ll\cb$ we can verify that the configuration
in \Eref{inftr} is stable w.r.t the excitations of the
non-inflaton fields, finding the expressions of the masses squared
$\what m^2_{\chi^\al}$ (with $\chi^\al=\theta$ and $s$) arranged
in \Tref{tab1}, which approach rather well the quite lengthy,
exact expressions taken into account in our numerical computation.
These expressions assist us to verify that the positivity of
$\what m^2_{\chi^\al}$ requires $1<N<6$ for $K=K_1$ and $N_S<6$
for $K=K_{2}$. Moreover, for both masses squared we obtain $\what
m^2_{\chi^\al}\gg\Hhi^2=\Vhi/3$ for $\sgf\leq\sg\leq\sgx$ -- where
$\sgx$ and $\sgf$ are the values of $\sg$ when $\ks=0.05/{\rm
Mpc}$ crosses the horizon of nMI and at its end correspondingly.
In \Tref{tab1} we display the masses of the corresponding fermions
too. The derived mass spectrum can be employed in order to find
the one-loop radiative corrections, $\dV$, to $\Vhi$.  The
resulting $\dV$ lets intact our inflationary outputs, provided
that the renormalization-group mass scale $\Lambda$, is determined
by requiring $\dV(\sgx)=0$ or $\dV(\sgf)=0$. The possible
dependence of our findings on the choice of $\Lambda$ can be
totally avoided if we confine ourselves to $0<N_S<6$  resulting to
$\Ld\simeq(1.1-2.5)\cdot10^{-3}$ for $K=K_1$ or
$\Ld\simeq(1.72-2.9)\cdot10^{-5}$ for $K=K_2$. Under these
circumstances, our inflationary predictions can be exclusively
reproduced by using $\Vhi$ in \Eref{Vhi} -- cf. \cref{jhep}.



\section{Inflation Analysis}\label{inf}

From the setting of our models we can easily deduce that their
free parameters, for fixed $n$, are $\rs=\cb/\ca^2$ and
$\ld/\ca^2$ and not $\ca$, $\cb$ and $\ld$ as naively expected. In
fact, if we perform a rescaling $\sg=\tilde\sg/\ca$,
\Eref{actionJ} preserves its form replacing $\sg$ with $\tilde\sg$
where $\fr$ and $\Vjhi$, respectively, read
\beq\label{frVrs} \fr=1+\tilde\sg+\rs\tilde\sg^{2}~~\mbox{and}~~
\Vjhi=\ld^2\tilde\sg^{4}/4\ca^{4},\eeq
which, indeed, depend only on $\rs$ and $\ld/\ca^{2}$. Note that
here we have the same number of parameters with those employed in
the models of \crefs{nMkin,var,jhep} and one parameter more than
those in \cref{flinear}.

These parameters may be constrained by applying the inflationary
criteria. In particular, the period of slow-roll nMI is determined
in the EF by the condition \cite{review}
\beqs\beq{\ftn\sf
max}\{\eph(\phi),|\ith(\phi)|\}\leq1,\label{srcon}\eeq where the
slow-roll parameters $\eph$ and $\ith$ read
\beq\label{sr}\widehat\epsilon= \left({\Ve_{\rm
CI,\se}/\sqrt{2}\Ve_{\rm CI}}\right)^2
\>\>\>\mbox{and}\>\>\>\>\>\widehat\eta={\Ve_{\rm
CI,\se\se}/\Ve_{\rm CI}} \eeq\eeqs
and can be derived by employing $\Vhi$ in \Eref{Vhi} and $J$ in
\Eref{Jg}, without express explicitly $\Vhi$ in terms of $\se$. In
the limit $\ca\ll\cb$, the numerator in Eq.~(26) is dominated by
$4N\cb^2\sg^2$ whereas the denominator by $2\cb^2\sg^4$ and so we
can achieve the approximate formula
\beq  J\simeq\sqrt{2N}/\sg, \label{Jappr}\eeq
which turns out to be rather accurate. Inserting this into
\Eref{sr} we arrive at the following results
\beqs\beq\label{sr1} \sqrt{\eph}=\frac1{\sqrt{N}\fr}\Big( 2n \cb
\sg^2- (1 -n)\ca\sg-2 \Big)\eeq and
\begin{equation}\begin{aligned}\label{sr2} \ith&= \frac1{N\fr^2}\Bigg(8+\ca \sg\lf7 - 9 n + (n (8
n-9)-1)\cb\sg^2\rg\\ &+ \sg^2 \lf2 \ca^2 (1 -n)^2 + 4 \cb (n (2
\cb n \sg^2-5 )-1)\rg\Bigg)\,. \end{aligned}\end{equation}\eeqs We
can numerically verify that \Eref{srcon} is saturated for
$\sg=\sgf\ll1$, which is found from the condition
\beq \ith\lf\sgf\rg\simeq1~~\Rightarrow~~\sgf\simeq\frac{1 + 9
n}{N\ca \rs}\,, \label{sgf}\eeq
where we keep from the expression in \Eref{sr2} the most
significant terms for $\sg\ll1$.

The number of e-foldings $\Ns$ that the scale $\ks=0.05/{\rm Mpc}$
experiences during this nMI and the amplitude $\As$ of the power
spectrum of the curvature perturbations generated by $\sg$ can be
computed using the standard formulae
\begin{equation}
\label{Nhi}  \mbox{\sf\small (a)}~~\Ns=\int_{\sef}^{\sex}
d\se\frac{\Vhi}{\Ve_{\rm CI,\se}}~~~\mbox{and}~~~\mbox{\sf\small
(b)}~~\As= \left.\frac{1}{12\pi^2}\frac{\Ve_{\rm
CI}^{3}}{\Ve^2_{\rm CI,\se}}\right|_{\se=\sex},\eeq
where $\sgx~[\sex]$ is the value of $\sg~[\se]$ when $\ks$ crosses
the inflationary horizon. Taking into account \Eref{Jappr} and
$\sgx\gg\sgf$, from \sEref{Nhi}{a} we find
\beqs\begin{equation} \label{Nhia} \Ns\simeq-\frac{N(1 + n)}{2n(1
- n)}\ln\lf1- \frac{2n\ca\rs\sgx}{1-n}\rg\,. \end{equation}
Solving the equation above w.r.t $\sgx$ we find
\beq\label{s*}\sgx\simeq
\frac{(1-\ern)(1-n)}{2n\rs\ca}~~\mbox{where}~~\ern =
e^{-2(1-n)n\Ns/N(1+n)}\,.\eeq
Taking the limit $n\to0$ of the results above, we obtain
\beq \label{s0*}\Ns\simeq N\ca\rs\sgx~\Rightarrow~\sgx\simeq
\Ns/N\ca\rs.\eeq\eeqs
From the last expressions above we easily infer that there is a
lower bound on $\ca$, -- e.g. for $n=0$ we find
$\ca\gtrsim\Ns/N\rs$ -- above which $\sgx\leq1$ and so, our
proposal can be stabilized against corrections from higher order
terms.

Inserting \Eref{s*} into \sEref{Nhi}{b} we can derive a constraint
on $\ld/\ca^2$ for chosen $\rs$, i.e.,
\beqs\bea \label{lan}\frac{\ld}{\ca^2}&\simeq& 2^{\frac72 - 2 n}
\pi\ern\sqrt{\frac{3\As}{N}}\lf1+\ern+\frac{1-\ern}{n}\rg^n\nonumber
\\ &&(1- n)^n\lf\frac{n\rs}{1-\ern}\rg^{1 - n}\,.\eea
Substituting $\ern$ from its definition in \Eref{s*} and computing
the limit $n\to 0$, the above expression may be simplified as
\beq \label{lan0}\ld\simeq4
\sqrt{6N\As}\ca^2\pi\rs/\Ns\,.\eeq\eeqs Taking into account the
definition of $\rs$, we infer that $\ld$ is proportional to $\cb$
for $n=0$ similarly to the original nMI \cite{old,sm1}, where
$\ld$ is proportional to $\cat$.

The remaining inflationary observables are found from the
relations
\beqs\bea \label{ns} && \ns=\: 1-6\widehat\epsilon_\star\ +\
2\widehat\eta_\star,~~r=16\widehat\epsilon_\star, \\ \label{as} &&
\as =\:2\left(4\widehat\eta_\star^2-(n_{\rm
s}-1)^2\right)/3-2\widehat\xi_\star, \eea\eeqs
where the variables with subscript $\star$ are evaluated at
$\sg=\sgx$ and $\widehat\xi={\Ve_{\rm CI,\widehat\phi} \Ve_{\rm
CI,\widehat\phi\widehat\phi\widehat\phi}/\Ve^2_{\rm CI}}$.
Inserting $\sgx$ from \Eref{s*} into \eqs{sr1}{sr2} and then into
equations above we can obtain some analytical estimates. These
become more meaningful, expanding successively the results for low
$\rs, n$ and $1/\Ns$. Our final, quite accurate expressions are
\beqs\bea\ns &\simeq& 1 - \frac{2}{\Ns} - \frac{2 n}{N} - \frac{4
N}{\Ns^2} - \frac{16 n \rs}{\Ns},\label{ns1}\\ r &\simeq&
-\frac{32 n}{\Ns} + \frac{16 N}{\Ns^2} - \frac{32 n
N}{\Ns^2}(1+2\rs) \nonumber \\ &+& \frac{64 N^2}{\Ns^3}\lf1 +8n\rg
\rs,\label{r1} \\ \as &\simeq&-\frac{2}{\Ns^2} - \frac{4 n}{\Ns^2}
- \frac{12 N}{\Ns^3}.\label{as1}\eea\eeqs
Due to the approximations made, the results for $n=0$ are not
obtained by taking the relevant limit of the expressions above.
Repeating the procedure, i.e., plugging \Eref{s0*} into
\eqs{sr1}{sr2} and expanding successively the results for low
$\rs$ and $1/\Ns$ we find
\beqs\bea \ns&\simeq& 1 - \frac{2}{\Ns} + \frac{2 N}{\Ns^2} -
\frac{8N}{\Ns^2}\rs, \label{ns0}\\ r &\simeq& \frac{16 N}{\Ns^2} -
\frac{32 N^2}{\Ns^3} + \frac{64 N^2}{\Ns^3}\rs,\label{r0} \\ \as
&\simeq& -\frac{2}{\Ns^2} + \frac{6 N}{\Ns^3} -
\frac{20N}{\Ns^3}\rs.\label{as0}\eea\eeqs
We remark a weak dependence of the results on $n$ and $\rs$ which
may deviate from the ones obtained in the contemporary nMI
\cite{old, sm1}.

\section{Numerical Results}\label{res}

Our analytic findings above can be verified numerically and
employed in order to delineate the available parameter space of
the models. In particular, we confront the quantities in
\Eref{Nhi} with the observational requirements \cite{flinear}
\beqs\bel&\Ns\simeq61.3+\frac12\ln\lf{\Vci(\sgx)\fr(\sgx)\over
g_{\rm rh*}^{1/6}\Vci(\sgf)^{1/2}}\rg\simeq58-60\label{Ntot}\\
&\mbox{and}~~~\As\simeq2.105\cdot10^{-9},~~~~~~\label{Prob}\end{align}\eeqs
where we assume that nMI is followed in turn by a oscillatory
phase, with mean equation-of-state parameter $w_{\rm
rh}\simeq1/3$, radiation and matter domination. Also $g_{\rm
rh*}=228.75$ or $106.75$ is the energy-density effective number of
degrees of freedom which corresponds to the Minimal SUSY SM or SM
spectrum respectively.

Enforcing \eqs{Ntot}{Prob} we can restrict $\ld/\ca^{2}$ and
$\sgx$ and compute the models' predictions via \eqs{ns}{as}, for
any selected $\rs$ and $n$. The outputs, encoded as lines in the
$\ns-\rw$ plane, are compared against the observational data
\cite{plcp,gwsnew} in \Fref{fig1} for $K=K_1$ -- here
$\rw=16\eph(\sg_{0.002})$ where $\sg_{0.002}$ is the value of
$\sg$ when the scale $k=0.002/{\rm Mpc}$, which undergoes $\what
N_{0.002}=(\Ns+3.22)$ e-foldings during nMI, crosses the horizon
of nMI. We draw dot-dashed, double dot-dashed, solid, dotted and
dashed lines for $n=0.001, 0.005, 0, -0.005$ and $-0.01$
respectively and show the variation of $\rs$ along each line. We
take into account the data from \plk\ and \emph{Baryon Acoustic
Oscillations} ({\sf\ftn BAO}) and the {\sf\ftn BK14} data taken by
the \bcp\ CMB polarization experiments up to and including the
2014 observing season. Fitting the data above \cite{plin,gwsnew}
with $\Lambda$CDM$+r$ we obtain the marginalized joint $68\%$
[$95\%$] regions depicted by the dark [light] shaded contours in
\Fref{fig1}. Approximately we may write
\begin{equation}  \label{data}
\mbox{\small\sf
(a)}~\ns=0.967\pm0.0074~~\mbox{and}~~\mbox{\small\sf
(b)}~r\leq0.07,
\end{equation}
at 95$\%$ \emph{confidence level} ({\sf\ftn c.l.}) with
$|\as|\ll0.01$. The constraint on $|\as|$ is readily satisfied
within the whole parameter space of our models.

\begin{figure}[!t]
\centering
\includegraphics[width=60mm,angle=-90]{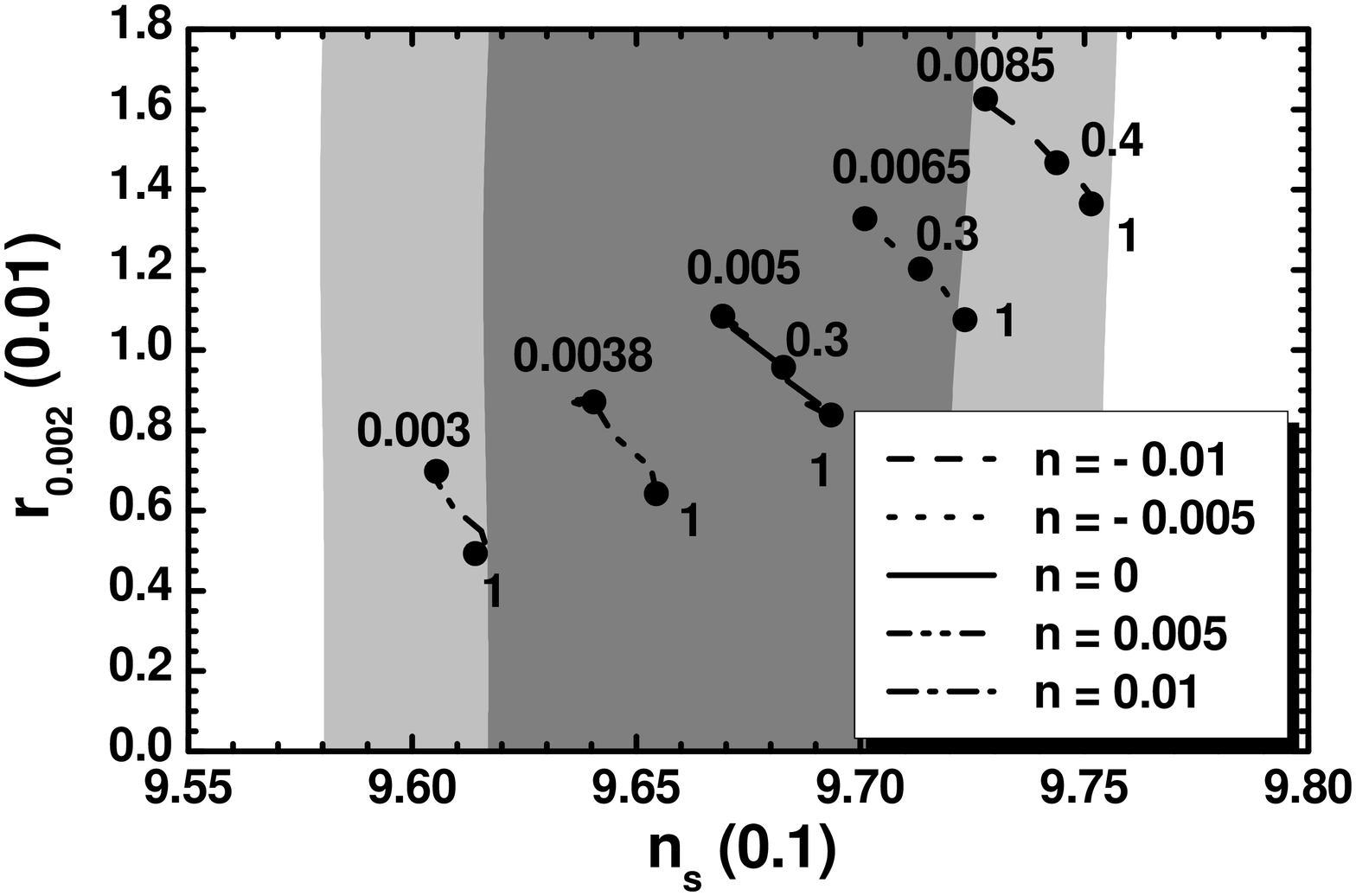}\\
\begin{ruledtabular}
\begin{tabular}{c||c|c||c|c|c|c}
$n/$&\multicolumn{2}{c||}{$r^{\rm
min}/0.01$}&\multicolumn{2}{c|}{$r_{21}^{\rm min}/0.001$}&\multicolumn{2}{c}{$r^{\rm max}/0.01$}\\
\cline{2-7}$0.01$& $K=K_1$&$K=K_2$&$K=K_1$&$K=K_2$&$K=K_1$&$K=K_2$
\\\colrule
$-1$&$1.5$&$1.3$&$8.5$&$22$&$1.7$&$1.4$\\
$-0.5$&$1.2$&$0.9$&$6.5$&$14$&$1.4$&$1.1$\\
$0$&$0.9$&$0.66$&$5$&$9$&$1.2$&$0.78$\\
$0.5$&$0.74$&$0.46$&$3.8$&$5.9$&$0.95$&$0.58$\\
$1$&$0.55$&$0.32$&$3$&$5$&$0.77$&$0.4$\\
\end{tabular}
\end{ruledtabular}
\caption{\sl \small Allowed curves in the $\ns-\rw$ plane for
$K=K_1$, $n=0,\pm0.005,\pm0.01$ (using the line types shown in the
plot legend) and various $\rs$'s indicated on the lines. The
marginalized joint $68\%$ [$95\%$] regions from \plk, {\sf\ftn
BK14} and  BAO data \cite{plin} are depicted by the dark [light]
shaded contours. The minimal and maximum $r$'s (corresponding to
the minimal $\rs$'s) for the $n$'s shown in the plot are shown in
the table for $K=K_1$ or $K_2$.}\label{fig1}
\end{figure}

From \Fref{fig1} we observe that the allowed $\ns$ and $r$ values
increase as $n$ decreases. More interestingly, for any selected
$n$ there is a lower ($\rs^{\rm min}$) and an upper ($\rs^{\rm
max}$) bound on $\rs$ which is translated correspondingly to an
upper ($r^{\rm max}$) and a lower ($r^{\rm min}$) bound on $r$.
Namely, the origin of $\rs^{\rm min}$ comes from the requirement
that $\ld$ has to remain within the domain of the validity of the
perturbation theory and so it has to be lower than about
$\sqrt{4\pi}\simeq3.5$. On the other limit, the various lines
terminate for $\rs^{\rm max}\simeq1$, beyond which the effective
theory ceases to be well defined -- see \Sref{uv}. The bounds
provided by these constraints together with the $\rs^{\rm min}$'s
are listed in the Table of \Fref{fig1}. These values are found not
only for $K=K_1$ but also for $K=K_2$ for the sake of comparison.
Indeed, had we employed $K=K_2$ the various lines in \Fref{fig1}
would have been remained almost intact with the same $\rs^{\rm
max}$'s and the $\rs^{\rm min}$'s acquiring the values arranged in
the Table. From our findings we see that, for $K=K_1$, a little
larger $r$'s are achieved, in accordance with our analytic
expressions in \eqs{r1}{r0}. Since $r\gtrsim0.0032$, our models
are testable by the forthcoming experiments \cite{bcp3}, which are
expected to measure $r$ with an accuracy of $10^{-3}$.

For $n=0$ and $N=3$ -- recall \Eref{kp} -- we obtain the results
for the non-SUSY regime. Moreover, this $n$ value results to
integer $N$'s, -- via \Eref{ndef} -- in the SUSY regime, which may
be regarded as the theoretically most favored. In particular, for
$K=K_1$ and $N=3$ we get
\beq\label{res1a} 9.67\lesssim
\frac{\ns}{0.1}\lesssim9.69\>\>\mbox{and}\>\>1.2\gtrsim
\frac{r}{0.01}\gtrsim0.9\,,\eeq
whereas for $K=K_2$ and $N=2$ we get the same $\ns$ interval with
$r$ ranging between the two values indicated in the Table of
\Fref{fig1}. In both cases we have $|\as|\simeq0.00049-0.00054$.
Therefore, the compatibility of these outputs with the
observational values in \Eref{data} is certainly impressive.

\begin{figure}[!t]
\includegraphics[width=60mm,angle=-90]{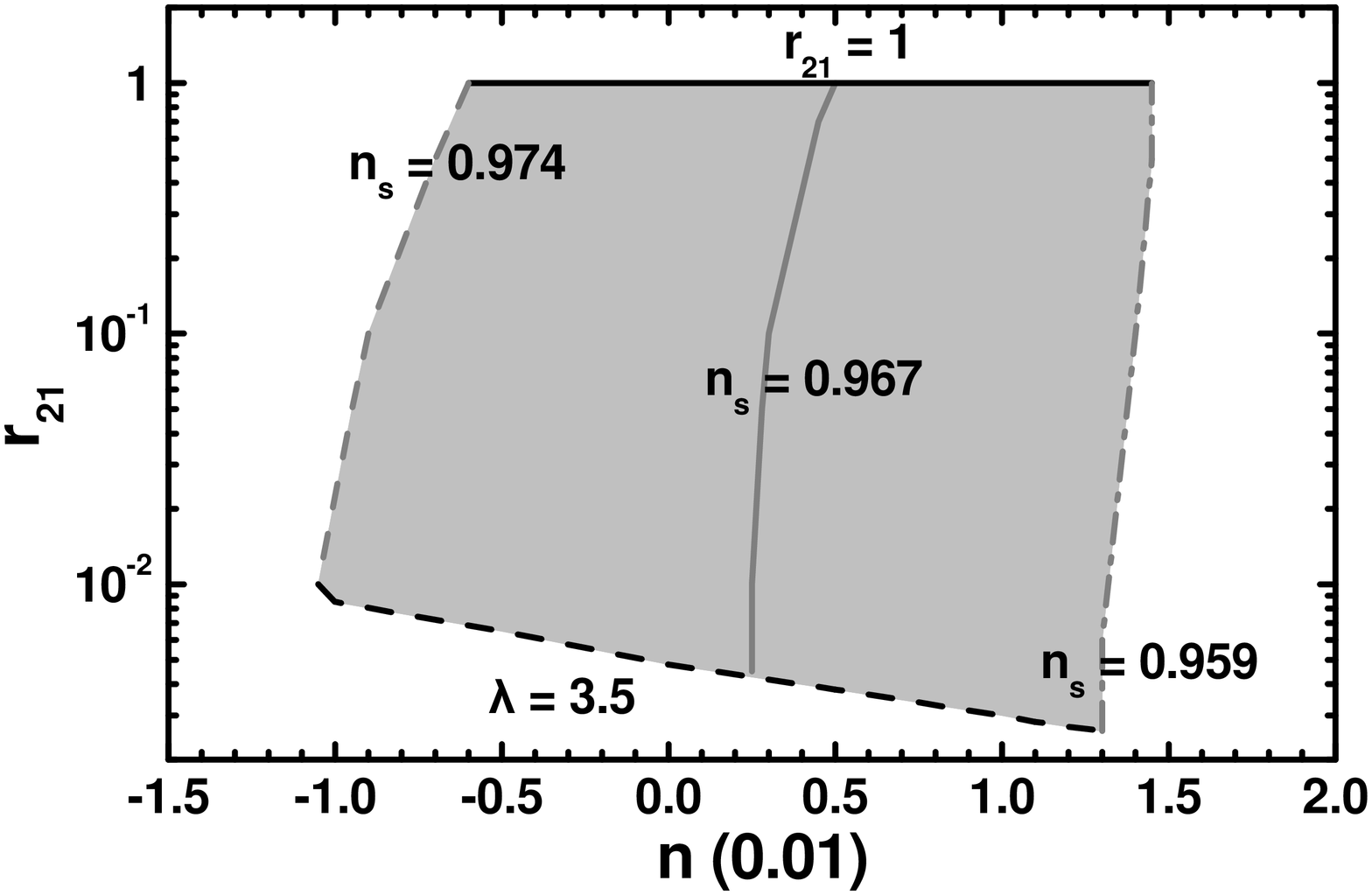}
\caption{\sl Allowed (shaded) region in the $n-\rs$ plane for
$K=K_{1}$. The constraint fulfilled along each line is also shown
on it.}\label{fig2}
\end{figure}

Varying continuously $n$ we can identify the allowed region in the
$n-\rs$ plane -- as in \Fref{fig2}. The allowed (shaded) region is
bounded by the solid black line, which corresponds to
$\rs\simeq1$, the dashed black line which originates from the
bound $\ld\leq3.5$ and the dot-dashed and dashed gray lines along
which the lower and upper bounds on $\ns$ in \Eref{data} are
saturated respectively. We remark that increasing $n$, with fixed
$\rs$, $\ns$ decreases, in accordance with our findings in
\Fref{fig1}. Fixing $\ns$ to its central value in \Eref{data}, we
obtain the gray solid line along which we get clear predictions
for $n$ and $r$. Namely,
\beqs\beq\label{res2a} 2\lesssim
\frac{n}{0.001}\lesssim5,~0.0046\lesssim
{\rs}\lesssim1~~\mbox{and}~~9.8\gtrsim
\frac{r}{0.001}\gtrsim6.4,\eeq
with $\as/10^{-4}\simeq-(4.5-4.9)$. Had we employed $K=K_{2}$, the
allowed region in Fig.~\ref{fig2} would have been remained very
similar whereas \Eref{res2a} would have been modified as follows
\beq\label{res2b} 0.3\lesssim
\frac{n}{0.001}\lesssim2,~0.0085\lesssim
{\rs}\lesssim1~\mbox{and}~~7.8\gtrsim
\frac{r}{0.001}\gtrsim5.6.\eeq\eeqs
 %
\begin{figure}[!t]
\includegraphics[width=60mm,angle=-90]{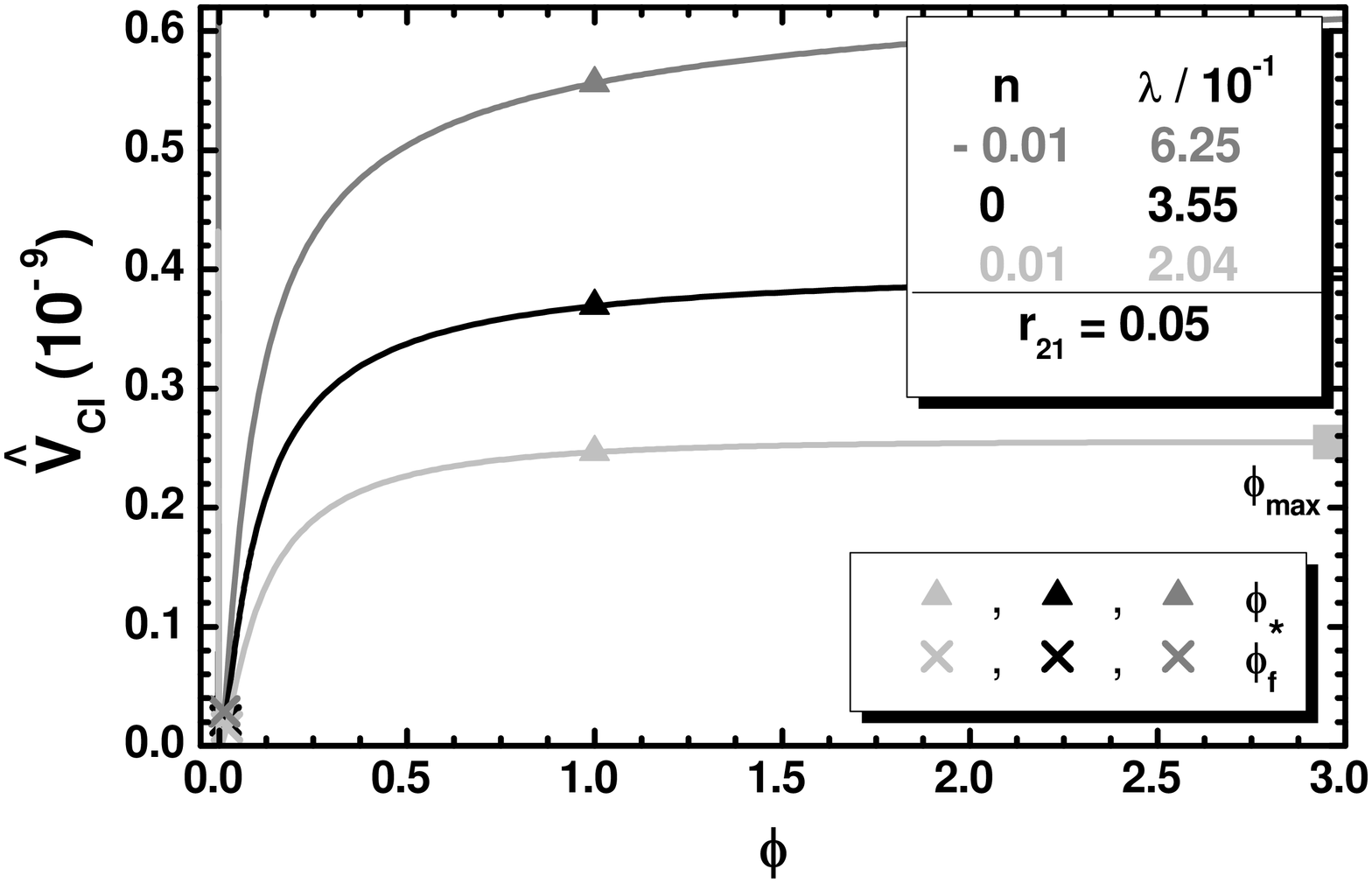}
\begin{ruledtabular}
\begin{tabular}{c||c|c|c|c||c|c|c|c}
$n/$&\multicolumn{4}{c||}{\sc Numerical
Values}&\multicolumn{4}{c}{\sc Analytic Values}\\\cline{2-9}
$0.01$&$\ld$&$\ca/10^2$&$\ns$&$r/0.01$&$\ld$&$\sgx$&$\ns$&$r/0.01$\\\colrule
&\multicolumn{8}{c}{$K=K_1$}\\\cline{2-9}
$-1$&$0.625$&$5.3$&$0.973$&$1.7$&$0.67$&$0.97$&$0.975$&$1.5$\\
$0$&$0.355$&$4.2$&$0.967$&$1.2$&$0.397$&$0.95$&$0.968$&$1.1$\\
$1$&$0.204$&$3.35$&$0.96$&$0.78$&$0.23$&$0.95$&$0.963$&$0.64$\\\colrule
&\multicolumn{8}{c}{$K=K_2$}\\\cline{2-9}
$-1$&$1.56$&$8.9$&$0.975$&$1.4$&$1.67$&$0.97$&$0.975$&$1.4$\\
$0$&$0.64$&$6.3$&$0.967$&$0.78$&$0.73$&$0.95$&$0.968$&$0.79$\\
$1$&$0.275$&$4.55$&$0.956$&$0.43$&$0.32$&$0.96$&$0.955$&$0.34$\\ 
\end{tabular}
\end{ruledtabular}
\caption{\sl \small The inflationary potential $\Vhi$ as a
function of $\sg$ for $K=K_1$, $\sg>-0.1$, $\rs=0.05$, and
$n=-0.01$, $\ld=0.625$ (gray line), $n=0$, $\ld=0.355$ (black
line), or $n=+0.01$, $\ld=0.204$ (light gray line). The values of
$\sgx$, $\sgf$ and $\sg_{\rm max}$ (for $n=0.01$) are also
indicated. Some of the parameters of our models for the $n$'s
shown in the plot are displayed in the Table for $K=K_1$ or $K_2$
and employing our numerical or analytic formulae.}\label{fig3}
\end{figure}

We complete our numerical analysis by studying the structure of
$\Vhi$. We fix $K=K_1$, $\sgx=1$ and $\rs=0.05$ and draw $\Vhi$ in
\Eref{Vhi} (gray, black and light gray lines) as a function of
$\sg$ for $n=-0.01, 0$ and $0.01$ respectively.  The corresponding
values of $\ld$, $\ca$, $\ns$ and $r$ are listed in the second,
third, fourth and fifth leftmost columns of the Table below the
graph, not only for $K=K_1$ but also for $K=K_2$ for comparison
purposes. In all cases $\as\simeq-5\cdot10^{-4}$. These results
are obtained by our numerical code taking into account exact
expressions for $\Vci, J$ and the other observables -- i.e.,
Eqs.~(\ref{Vhi}), (\ref{Jg}), (\ref{Nhi}), (\ref{ns}) and
(\ref{as}). These values are also consistent with those obtained
by employing the formulas of \Sref{inf} -- i.e.,
Eqs.~(\ref{Nhia}), (\ref{lan}) and (\ref{ns1}) -- (\ref{as0}) --
and displayed in the four rightmost columns of the Table in
\Fref{fig3}. Note that in the case of analytic expressions we
prefer to compare $\sgx$ derived by \Eref{s*} with $\sgx=1$, used
in all cases numerically, and let $\ca$ as input parameter.
Moreover, we observe that $\Vhi$ is a monotonically increasing
function of $\sg$  for $n\leq0$ whereas it develops a maximum at
$\sg_{\rm max}=2.96$, for $n=0.01$, which leads to a mild tuning
of the initial conditions of nMI since $\sgx\ll\sg_{\rm max}$. It
is also remarkable that $r$ increases with the inflationary scale,
$\Vhi^{1/4}$, which in all cases is roughly of the order of
$0.01\mP$. Since $\Vhi^{1/4}\ll\mP$ and $\mP$ is the UV cut-off
scale of the theory -- as we show in \Sref{uv} --, the classical
approximation, used in our analysis is perfectly valid. Finally,
it is worth emphasize that $\Vhi$ is of Starobinsky-type although
$\vev{\sg}=0$ in sharp contrast to the models of induced-gravity
inflation \cite{nIG,igHI,gian} where $\vev{\sg}\gg0$.

\section{Effective Cut-Off Scale}\label{uv}

The main motivation of the nMI proposed in this work is that it is
unitarity-safe, despite the fact that its implementation with
\sub\ $\phi$ values requires relatively large $\ca$ and $\cb$
values -- see, e.g., the Table of \Fref{fig1}. To show that this
fact is valid we extract below the UV cut-off scale, $\Qef$,
expanding the action in the JF -- see \Sref{uv1} -- and in the EF
-- see \Sref{uv2}.

\subsection{\sf\ftn Jordan Frame Computation}\label{uv1}

Thanks to the special dependence of $\fr$ on $\phi$ in \Eref{fr},
the interaction between the excitation of $\phi$ about
$\vev{\phi}=0$, $\dph$, and the graviton, $h^{\mu\nu}$ preserves
the perturbative unitarity for $\rs\leq1$. Indeed, we expand
$g_{\mu\nu}$ about the flat spacetime metric $\eta_{\mu\nu}$ and
the inflaton $\phi$ about its v.e.v,
\beq
g_{\mu\nu}\simeq\eta_{\mu\nu}+h_{\mu\nu}\>\>\>\mbox{and}\>\>\>\phi=0
+\dph\,.\eeq
Retaining only the terms with up to two four-dimensional
derivatives of the excitations, the part of the lagrangian
corresponding to the two first terms in the r.h.s of
\Eref{actionJ} takes the form \cite{cutof, nIG}
\beq\begin{aligned} \delta{\cal L}&=-{\vev{\fr}\over8}{G}_{\rm
EH}\lf h^{\mu\nu}\rg +\frac12\vev{f_{\rm K}}\partial_\mu
\dph\partial^\mu\dph\\&+\frac12G_{\cal R}\lf h^{\mu\nu}\rg\lf
\vev{f_{\cal R,\phi}}\dph+\frac12 \vev{f_{\cal R,\phi\phi}}\dph^2\rg\\
&=-{1\over8}G_{\rm EH}\lf \bar h^{\mu\nu}\rg+ \frac12\partial_\mu
\overline\dph\partial^\mu\overline\dph+\frac1{\Qef}\overline\dph^2\,\Box
\bar h\, +\ \cdots\label{L2}\end{aligned}\eeq
where $\vev{f_{\rm K}}=1$ for the non-SUSY case and $\vev{f_{\rm
K}}=1+(N-3)\ca^2/2$ -- see \Eref{omab} -- for our SUGRA scenaria.
Therefore, for $N=3$ the results below reduce to that obtained in
the non-SUSY regime. The functions $G_{\rm EH}$ and $G_{\cal R}$
are identical to the functions $F_{\rm EH}$ and $F_{\cal R}$
defined in \cref{nIG}; $\bar h_{\mu\nu}$ and $\overline\dph$ are
the JF canonically normalized fields defined by the relations
\bel \nonumber &\overline\dph=\sqrt{\frac{\vev{\bar f_{\cal
R}}}{\vev{\fr}}}\dph\>\>\>\mbox{with}\>\>\>\bar f_{\cal R}=f_{\rm
K}\fr+\frac32 f_{\cal R,\phi}^2\\&\mbox{and}\>\>\> \bar
h_{\mu\nu}= \sqrt{\vev{\fr}}\,h_{\mu\nu}+\frac{\vev{f_{\cal
R,\phi}}}{\sqrt{\vev{\fr}}}\eta_{\mu\nu}\dph\,. \label{Jcan}
\end{align} Taking into account that $\vev{\sg}=0$ we find
$\vev{\fr}=1$ and $\vev{\bar f_{\cal R}}=1+N\ca^2/2\simeq
N\ca^2/2$.

The possible problematic process, which causes \cite{cutoff}
concerns about the unitarity-violation, is the
$\overline\dph-\overline\dph$ scattering process via $s$-channel
graviton, $\overline{h}^{\mu\nu}$, exchange originating from the
last term in the r.h.s of \Eref{L2} with
$\overline{h}=\overline{h}^\mu_{\mu}$. The UV cut-off scale $\Qef$
is identified as follows
\beq \Qef^{-1}\simeq\cb\frac{\sqrt{\vev{\fr}}}{\vev{\bar f_{\cal
R}}}\simeq\frac{2\rs}{N}~\Rightarrow~\Qef\sim\rs^{-1}\,.\eeq
Therefore, the theory retains the perturbative unitarity up-to
$\mP$ for $\rs\leq1$.

\subsection{\sf\ftn Einstein Frame Computation}\label{uv2}

Alternatively, $\Qef$ can be determined in EF, following the
systematic approach of \cref{riotto}. We concentrate here on the
SUGRA versions of our model. The transition to the non-SUSY case
can be easily achieved setting $n=0$ or $N=3$. The EF (canonically
normalized) inflaton is
\beq\dphi=\vev{J}\dph\>\>\>\mbox{with}\>\>\>\vev{J}=\sqrt{1+\ca^2\frac{N}{2}}\simeq\ca\sqrt{\frac{N}{2}}\,.
\label{dphi} \eeq
From the last expression, we can clearly appreciate the importance
of the linear term in $\fr$, \Eref{fr}, to distinguish $\dphi$
from $\dph$ -- recall that in the standard non-minimal Higgs
inflation \cite{cutoff,cutof} $\dphi=\dph$. As anticipated in
\Sref{intro}, this fact implies that our models are valid up to
$\mP$. To prove it, we focus on the second term in the r.h.s of
\Eref{action1} for $\mu=\nu=0$ and we expand it about
$\vev{\phi}=0$ in terms of $\dphi$. Our result is written as
\beqs\bel J^2 \dot\phi^2&\simeq\lf1 - 2(1-2\rs) \sqrt{\frac2N}
\dphi + \lf3-10\rs\rg\frac{2}{N}\dphi^2\nonumber \right.\\&-\left.
\lf2-9\rs\rg\frac4{N}\sqrt{\frac2N}\dphi^3
+\cdots\rg\dot\dphi^2\,,\label{Jexp}\end{align}
where we neglect terms suppressed by powers of $\rs$ and inverse
powers of $\ca$ -- since $\rs\leq1$ and $\ca\gg1$. Expanding
similarly $\Vhi$, see \Eref{Vhi}, in terms of $\dphi$ we have
\beq\begin{aligned}\Vhi&\simeq\frac{\ld^2\dphi^4}{N^2\ca^4} \lf
1-2(1+n) \sqrt{\frac{2}{N}}\dphi\right.\\&
+\left.\Big(3+5n-2(1+n)\rs\Big)\frac{2}{N}\dphi^2\right.\\&
-\left.\lf2+\frac{13}{3}n-(3+5n)\rs\rg\frac4{N}\sqrt{\frac2N}\dphi^3
+\cdots\rg. \label{Vexp}\end{aligned}\eeq\eeqs
Consequently, we verify again that our models preserve the
perturbative unitarity up to $\mP$ for $\rs\leq1$.


\section{Conclusions and Perspectives} \label{con}

We presented a unitarized version of non-minimal inflation (i.e.
nMI) which fits the \plk\ data very well. The main novelty of our
proposal is the consideration of a linear term into the frame
function, \Eref{fr}, -- involving the parameter $\ca$ -- apart
from the usual quadratic term proportional to $\cb$ and the
quartic potential in \Eref{Vn}. This setting can be elegantly
implemented not only in non-SUSY regime but also within SUGRA,
employing the super- and \Ka s given in \eqs{Whi}{K1} or
(\ref{K2}) and extending the parameter space of the model by one
parameter $n$ defined in \Eref{ndef}. Our investigation reveals
that $n$ has to be tuned into the interval $\lf(-0.01)-0.013\rg$.
Confining ourselves to the most natural value $n=0$, we achieved
observational predictions which may be tested in the near future
and converge towards the ``sweet'' spot of the present data for
$\rs=\cb/\ca^2$ into the range $(4.6\cdot10^{-3}-1)$ -- see
\Fref{fig1}. Thanks to the presence of the non-vanishing $\ca$, no
problem with the perturbative unitarity arises for $\rs\leq1$,
although the attainment of nMI with subplanckian values requires
relatively large $\ca$'s (and $\cb$'s). It is gratifying, finally,
that the allowed parameter space of our models can be studied
analytically and rather accurately.

As a last remark, we would like to point out that, although we
have restricted our discussion to a gauge singlet inflaton, the
applicability of our proposal can be easily extended to gauge
non-singlet fields. Indeed, the unitarization of non-minimal Higgs
inflation based on the potential $V_{\rm
HI}=\ld^2(\phc^\dagger\phc)^2/4$ -- where $\phc$ is now a Higgs
field in the fundamental representation of an $SU({\cal N})$ gauge
group --, according to our suggestion here, requires the
consideration of the frame function
$\fr=1+\ca\sqrt{\phc^\dagger\phc}+\cb\phc^\dagger\phc$ -- cf.
\cref{sm1,leel}. The second non-analytic term in the r.h.s of the
expression above, although unusual, is perfectly acceptable. In
this case, the inflationary predictions are expected to be quite
similar to the ones obtained here, although the parameter space
may be further restricted from the data on the Higgs mass. Indeed,
we should take into account the renormalization-group running of
the various parameters from the inflationary up to the electroweak
scale in order to connect convincingly the high- with the
low-energy phenomenology -- cf. \cref{hmass}. Since our main aim
here is the demonstration of the modification on the observables
of nMI due to the introduction of the linear term in $\fr$, we
opted to utilize just a gauge-singlet inflaton.

\paragraph*{\small\bfseries\scshape Acknowledgment} {\small I would like to acknowledge
Jos\'e Ra-m\'on Espinosa  for a useful correspondence.}

\vspace*{-.55cm}

\def\ijmp#1#2#3{{\sl Int. Jour. Mod. Phys.}
{\bf #1},~#3~(#2)}
\def\plb#1#2#3{{\sl Phys. Lett. B }{\bf #1}, #3 (#2)}
\def\prl#1#2#3{{\sl Phys. Rev. Lett.}
{\bf #1},~#3~(#2)}
\def\rmp#1#2#3{{Rev. Mod. Phys.}
{\bf #1},~#3~(#2)}
\def\prep#1#2#3{{\sl Phys. Rep. }{\bf #1}, #3 (#2)}
\def\prd#1#2#3{{\sl Phys. Rev. D }{\bf #1}, #3 (#2)}
\def\npb#1#2#3{{\sl Nucl. Phys. }{\bf B#1}, #3 (#2)}
\def\npps#1#2#3{{Nucl. Phys. B (Proc. Sup.)}
{\bf #1},~#3~(#2)}
\def\mpl#1#2#3{{Mod. Phys. Lett.}
{\bf #1},~#3~(#2)}
\def\jetp#1#2#3{{JETP Lett. }{\bf #1}, #3 (#2)}
\def\app#1#2#3{{Acta Phys. Polon.}
{\bf #1},~#3~(#2)}
\def\ptp#1#2#3{{Prog. Theor. Phys.}
{\bf #1},~#3~(#2)}
\def\n#1#2#3{{Nature }{\bf #1},~#3~(#2)}
\def\apj#1#2#3{{Astrophys. J.}
{\bf #1},~#3~(#2)}
\def\mnras#1#2#3{{MNRAS }{\bf #1},~#3~(#2)}
\def\grg#1#2#3{{Gen. Rel. Grav.}
{\bf #1},~#3~(#2)}
\def\s#1#2#3{{Science }{\bf #1},~#3~(#2)}
\def\ibid#1#2#3{{\it ibid. }{\bf #1},~#3~(#2)}
\def\cpc#1#2#3{{Comput. Phys. Commun.}
{\bf #1},~#3~(#2)}
\def\astp#1#2#3{{Astropart. Phys.}
{\bf #1},~#3~(#2)}
\def\epjc#1#2#3{{Eur. Phys. J. C}
{\bf #1},~#3~(#2)}
\def\jhep#1#2#3{{\sl J. High Energy Phys.}
{\bf #1}, #3 (#2)}
\newcommand\jcap[3]{{\sl J.\ Cosmol.\ Astropart.\ Phys.\ }{\bf #1}, #3 (#2)}
\newcommand\njp[3]{{\sl New.\ J.\ Phys.\ }{\bf #1}, #3 (#2)}
\def\prdn#1#2#3#4{{\sl Phys. Rev. D }{\bf #1}, no. #4, #3 (#2)}
\def\jcapn#1#2#3#4{{\sl J. Cosmol. Astropart.
Phys. }{\bf #1}, no. #4, #3 (#2)}
\def\epjcn#1#2#3#4{{\sl Eur. Phys. J. C }{\bf #1}, no. #4, #3 (#2)}


\begin{thebibliography}{99}
\section*{\refname} \ignorespaces

\bibitem{plcp} N.~Aghanim {\it et al.} [\plk\ Collaboration], \arxiv{1807.06 209}.

\bibitem{plin}   Y.~Akrami {\it et al.} [\plk\ Collaboration],
\arxiv{1807.06211}.


\bibitem{gwsnew}  P.A.R.~Ade {\it et al.} [{\sc BICEP2}/{\it Keck Array}
Collaborations],  {\sl Phys.\ Rev.\ Lett.} {\bf 116}, 031302
(2016) [\arxiv{1510.09217}].


\bibitem{old} D.S. Salopek, J.R. Bond and J.M. Bardeen,
{\sl Phys. Rev. D }{\bf 40}, 1753 (1989); J.L.~Cervantes-Cota and
H.~Dehnen, \prd{51}{1995}{395} [\astroph{9412032}].

\bibitem{sm1} J.L.~Cervantes-Cota and H.~Dehnen, \npb{442}{1995}{391}
[\astroph{9505069}]; F.L.~Bezrukov and M.~Shaposhnikov,
\plb{659}{2008}{703} [\arxiv{0710.3755}].





\bibitem{cutoff} J.L.F.~Barbon and J.R.~Espinosa,
\prd{79}{2009}{081302} [\arxiv{0903.0355}]; C.P.~Burgess,
H.M.~Lee, and M.~Trott, \jhep{07}{2010}{007} [\arxiv{1002. 2730}].

\bibitem{riotto} A.~Kehagias, A.M.~Dizgah, and A.~Riotto, \prd{89}{2014}{043527}
[\arxiv{1312.1155}].

\bibitem{cutof} F.~Bezrukov \etal, \jhep{016}{2011}{01}
[\arxiv{1008.5157}].


\bibitem{gianlee} G.F. Giudice and H.M. Lee, \plb{694}{2011}{294}
[\arxiv{1010.1417}].


\bibitem{john} R.N.~Lerner and J.~McDonald, \prd{82}{2010}{103525}
[\arxiv{1005.2978}].



\bibitem{lee} H.M.~Lee, {\sl Eur.\ Phys.\ J.\ C }{\bf 74}, 3022 (2014)
[\arxiv{1403. 5602}].



\bibitem{jhep} G.~Lazarides and C. Pallis, {\sl J. High Energy Phys.} {\bf 11},
114 (2015) [\arxiv{1508.06682}].

\bibitem{nMkin} C.~Pallis,~\prdn{91}{2015}{123508}{12} [\arxiv{ 1503.05887}];
C.~Pallis, {\sl Phys. Rev. D} {\bf 92}, no. 12, 121305(R) (2015)
[\arxiv{1511.01456}].


\bibitem{var} C.~Pallis, \jcapn{10}{2016}{037}{10} [\arxiv{1606.09607}].

\bibitem{tokareva}   Y.~Ema, {\sl Phys.\ Lett.\ B }{\bf 770}, 403 (2017) [\arxiv{1701.07665}];
D.~Gorbunov and A.~Tokareva, \arxiv{1807.02392}.

\bibitem{gian} G.F. Giudice and H.M. Lee, \plb{733}{2014}{58}
[\arxiv{1402.2129}].


\bibitem{R2r} C.~Pallis, \jcap{04}{2014}{024}; {\bf 07}, {01(E)} {(2017)}
[\arxiv{1312.3623}].

\bibitem{nIG} C.~Pallis, \jcap{08}{2014}{057} [\arxiv{ 1403.5486}]; C.~Pallis,
\jcap{10}{2014}{058} [\arxiv{1407.8522}].

\bibitem{igHI}  C.~Pallis and Q. Shafi,  {\sl Eur.\ Phys.\ J.\ C }{\bf 78}, no. 6, 523 (2018)
[\arxiv{1803.00349}].



\bibitem{jose} J.L.~F.~Barbon, J.A.~Casas, J.~Elias-Miro and J.R.~Espinosa,
{\sl J. High Energy Phys.} {\bf 09}, 027 (2015)
[\arxiv{1501.02231}].


\bibitem{leel} H.M.~Lee, {\sl Phys.\ Rev.\ D} {\bf 98}, no. 1, 015020 (2018)
[\arxiv{ 1802.06174}].


\bibitem{flinear} C. Pallis, {\sl Eur. Phys. J. C} {\bf 78}, no. 12, 1014 (2018) [\arxiv{18
07.01154}].

\bibitem{quad} C.~Pallis and Q.~Shafi, \jcap{03}{2015}{no.
03, 023} [\arxiv{1412.3757}].

\bibitem{linde1} M.B.~Einhorn and D.R.T.~Jones,
\jhep{03}{2010}{026} [\arxiv{0912.2718}]; H.M.~Lee,
\jcap{08}{2010}{003} [\arxiv{1005.2735}]; S.~Ferrara \etal,
\prd{83}{2011}{025008} [\arxiv{1008.2942}]; C.~Pallis and
N.~Toumbas, \jcap{02}{2011}{019} [\arxiv{1101.0325}].

\bibitem{su11} C.~Pallis and N.~Toumbas, \jcap{05}{2016}{no. 05, 015}
[\arxiv{1512.05657}];  C.~Pallis and N.~Toumbas, {\sl  Adv.\ High
Energy Phys.\ }{\bf 2017}, 6759267 (2017) [\arxiv{1612.09202}];
C.~Pallis, {\sl PoS EPS-HEP} {\bf 2017}, 047 (2017)
[\arxiv{1710.04641}].

\bibitem{roest} R.~Kallosh, A.~Linde and D.~Roest,
\jhep{11}{2013}{198} [\arxiv{1311.0472}]; R.~Kallosh, A.~Linde and
D.~Roest, \jhep{08}{2014}{052} [\arxiv{ 1405.3646}].

\bibitem{lofti} L. Boubekeur and D. Lyth, {\sl J. Cosmol. Astropart.
Phys. }{\bf 07}, 010 (2005) [\hepph{0502047}].

\bibitem{review} D.H.~Lyth and
A.~Riotto, {\sl Phys.\ Rept.} {\bf 314}, 1 (1999) [{\tt\ftn
hep-ph/ 9807278}]; J.~Martin, C.~Ringeval and V.~Vennin, {\sl
Physics of the Dark Universe} {\bf 5-6}, 75 (2014)
[\arxiv{1303.3787}].


\bibitem{bcp3} W.L.K. Wu \etal, {\sl J.\ Low.\ Temp.\ Phys.\ }{\bf 184}, no. 3-4, 765 (2016) [\arxiv{1601.00125}];
P. Andre \etal\ [PRISM Collaboration], \arxiv{1306.2259}; T.
Matsumura \etal, {\sl J. Low. Temp. Phys. }{\bf 176}, 733 (2014)
[\arxiv{1311.2847}]; F. Finelli \etal\  [CORE Collaboration]
\arxiv{1612.08270}.


\bibitem{hmass} F.L.~Bezrukov, A.~Magnin and M.~Shaposhnikov,
{\sl Phys.\ Lett.\ B }{\bf 675}, 88 (2009) [\arxiv{0812.4950}];
A.O.~Barvinsky \etal, \jcap{12}{2009}{003} [\arxiv{0904.1698}].
\end{thebibliography}
\end{document}